\documentclass[useAMS,usenatbib]{mn2e}
\usepackage{graphicx}
\usepackage{natbib}

\newcommand{\etal}{{et~al.}}
\newcommand{\ie}{$i.e.$}
\newcommand{\lsim}{\,\lower2truept\hbox{${<\atop\hbox{\raise4truept\hbox{$\sim$}}}$}\,}
\newcommand{\gsim}{\,\lower2truept\hbox{${>\atop\hbox{\raise4truept\hbox{$\sim$}}}$}\,}
\newcommand{\beq}{\begin{equation}}
\newcommand{\eeq}{\end{equation}}
\newcommand{\COBE}{$COBE$-DMR}
\newcommand{\WMAP}{$WMAP$}
\newcommand{\fstica}{{\sc{FastICA}}}

\newcommand{\vect}[1]{{\mathbfit{#1}}}

\def\aa{{\sl Astron.\ \&\ Astrophys.\ }}

\def\apj{{\sl Astrophys.\ J.\ }}

\def\apjs{{\sl Astrophys.\ J.\ Supp.\ }}

\def\ieeespl{{\sl IEEE\ Signal\ Processing\ Lett.\ }}

\def\mnras{{\sl MNRAS\ }}

\begin{document}

\title[Angular power spectrum of the \fstica\ CMB component from BEAST data]
{Angular power spectrum of the \fstica\ CMB component from BEAST data}

\author[S. Donzelli  et al.]
{S. Donzelli$^{1,2}$\thanks{simona.donzelli@mi.infn.it}, 
D. Maino$^{1}$, M. Bersanelli$^{1}$, J. Childers$^{3}$, N. Figueiredo$^{4}$,  
\and P.M. Lubin$^{3}$, P.R. Meinhold$^{3}$, I.J. O'Dwyer$^{5}$, M.D. Seiffert$^{6}$, 
T. Villela$^{7}$, \and B.D. Wandelt$^{5}$, C.A. Wuensche$^{7}$ \\
$^{1}$ Dipartimento di Fisica, Universit\`a di Milano, 
Via Celoria 16, 20133, Milano, Italy.\\
$^{2}$ SISSA/ISAS, Astrophysics Sector, Via Beirut 4, 34014, Trieste, Italy.\\
$^{3}$ Physics Department, University of California, Santa Barbara, CA 93106.\\
$^{4}$ Universidade Federal de Itajub\'a, Departamento de F\'isica e
Qu\'imica, Caixa Postal 50, 37500-903, Itajub\'a, MG, Brazil.\\
$^{5}$ Astronomy Department, University of Illinois at Urbana-Champaign,
Urbana, IL 61801-3074.\\
$^{6}$ Jet Propulsion Laboratory, Oak Grove Drive, Pasadena, CA 91109.\\
$^{7}$ Instituto Nacional de Pesquisas Espaciais, Divis\~ao de Astrof\'isica,
Caixa Postal 515, 12245-970, S\~ao Jos\'e dos Campos, SP, Brazil.} 

\date{Received
**insert**; Accepted **insert**}

\pagerange{\pageref{firstpage}--\pageref{lastpage}}
\pubyear{2005}

\maketitle

\label{firstpage}

\begin{abstract}
We present the angular power spectrum of the CMB component
extracted with \fstica\ from the Background Emission Anisotropy
Scanning Telescope (BEAST) data.  
BEAST is a 2.2 meter off-axis
telescope with a focal plane comprising 8 elements at Q (38-45~GHz)
and Ka (26-36~GHz) bands. It operates from the UC White Mountain Research
Station at an altitude of 3800 meters. The BEAST CMB angular power spectrum
has been already calculated by O'Dwyer \etal\ using only the
Q band data. With two input channels \fstica\ returns two possible 
independent components. We found that one of these two   
has an unphysical spectral behaviour
while the other is 
a reasonable CMB component.
After a detailed calibration procedure based on Monte-Carlo (MC) simulations
we extracted the angular power spectrum for the identified CMB component and
found a very good agreement with the already published BEAST CMB angular 
power spectrum and with the $WMAP$ data.

\end{abstract}

\begin{keywords}
methods -- data analysis -- techniques: image processing -- cosmic 
microwave background.
\end{keywords}

\section{Introduction}
\label{intro}

The recent outstanding results from \WMAP\ satellite 
\citep{bennett_etal_2003}
have definitely put us into the era 
of precision cosmology with an accurate determination of the
CMB angular power spectrum up to $\ell\simeq 800$. In 
addition the DASI experiment \citep{leitch_etal_2004} has clearly reported
detection of $E$ mode CMB polarization.
The situation will improve even further with
the {\sc Planck} satellite, a third generation
CMB space mission which will map microwave emission
over the whole sky with an unprecedented combination of angular
resolution and sensitivity. In the meantime a pletora of
both ground-based and balloon borne experiments will produce 
accurate measurements (better than \WMAP) over limited sky regions.

Today one of the main limitations to the accuracy is 
the presence of other astrophysical sources between us and the  
Last Scattering Surface, which contribute to the measured signal.  
These foreground contaminants consist mainly of Galactic emission 
(synchrotron, free-free and dust emission), compact galactic and extragalactic
sources, and the Sunyaev-Zel'dovich effect from cluster of galaxies. 
The challenge is to identify and remove such foreground emissions
with high accuracy and reliability in order to obtain cleaned CMB maps. 
This is crucial in deriving precise cosmological information from the
CMB power spectrum.

Many works have been dedicated to component separation, and different 
algorithms have been proposed. Traditional separation techniques, from Wiener
filtering \citep{tegmark_efstathiou_1996,bouchet_etal_1999,prunet_etal_2001} to Maximum 
Entropy Method (MEM)
\citep{hobson_etal_1998,stolyarov_etal_2002}, have generally been employed. They 
achieve good results, but they require prior knowledge about the signals to be separated
($e.g.$ spatial templates and frequency dependance), whereas the avaible full-sky foreground priors 
actually are not completely reliable. 

Recently a blind 
separation approach  has been developed, which works without the need of
priors, except about the statistical features of the components. 
Indeed this technique, based on the Independent Component
Analysis (ICA) \citep{comon_1994}, exploits  the statistical independence of the sky signals. 
It was first implemented as a neural network \citep{baccigalupi_etal_2000}, and then
optimized in a fast algorithm, \fstica\ \citep{maino_etal_2002}, which was successfully 
tested on simulated sky maps similar to those that {\sc Planck} will produce. 
\fstica\ has shown good performance also when applied to real data from 
\COBE\ \citep{maino_etal_2003}, with results on CMB anisotropy 
and foreground contamination consistent with previous and independent analyses. 

In this work we apply \fstica\ to another real data set, from the 
Background Emission Anisotropy Scanning Telescope (BEAST). In Section~\ref{fastica} 
we briefly recall the main features of the \fstica\  approach and its assumptions.
In Section~\ref{beast}, after describing the BEAST instrument and the maps produced, 
 we explain the procedure followed to apply \fstica\ to BEAST data. The results obtained
are presented in Section~\ref{results}, and the CMB reconstruction quality, 
tested with Monte Carlo simulations,
is analysed in Section~\ref{MC}. Sections~\ref{normal} deals with the normalisation of the CMB signal 
extracted with \fstica. In Section~\ref{spectrum} we extract the \fstica\ CMB spectrum. Finally, a 
critical discussion and conclusions are presented in Section~\ref{conclusion}.

\section{Component Separation with \fstica}
\label{fastica}

Before describing the \fstica\ method, it is useful to recall briefly the data model  
\fstica\ refers to and the principal assumptions from which it derives, described
in detail in \citet{maino_etal_2002}.

Let us suppose that the sky radiation, as a function of direction
$\vect{r}$ and frequency $\nu$, is a superposition of $N$ different
signals $s_{j}(\vect{r},\nu)$ and that it is observed by an experiment with $M$ frequency channels 
whose beam pattern is $B(\vect{r},\nu)$. Let us further suppose that, 
for each signal, frequency and spatial dependence can
be factored into two separated terms, $f_{j}(\nu)$ and $\bar{s}_{j}(\vect{r})$ respectively,   
and that $B$     
is shift-invariant and frequency-independent. Then the data model  
can be written as:
\beq \label{data_model}
{\bf{x}}(\vect{r}) = {\bf{A}} {\bar{\bf{s}}}(\vect{r}) * B(\vect{r}) +
{\bmath{\epsilon}}(\vect{r}) = {\bf{A}} {\bf{s}}(\vect{r}) + {\bmath{\epsilon}}(\vect{r})\, ,
\eeq
where each component, $s_{j}$, of the vector ${\bf{s}}$ is the
corresponding source function convolved with the $B$ beam pattern.
The matrix ${\bf{A}}$ is the mixing matrix, which includes the frequency response, and 
${\bf{\epsilon}(\vect{r})}$ is the instrumental noise term. 

The \fstica\ algorithm obtains both
the mixing matrix ${\bf{A}}$ and the signals ${\bf{s}}$ from observed data
${\bf{x}}$ 
assuming that
\begin{itemize}
\item the signals ${\bf{s}}$ are independent random processes on the map domain;
\item all the signal, but at most one, have non-Gaussian distribution.
\end{itemize}
A detailed explanation of this strategy can be found in \citet{hyvarinen_oja_1997}
and
\citet{hyvarinen_1999}, while its application in an astrophysical context 
is described in \citet{maino_etal_2002}.
Independent components are extracted maximising a suitable measure of non-Gaussianity
that is robust against noise: this is the the so-called neg-entropy. 

\fstica\ estimates separation matrix ${\bf{W}}$ row by row, 
maximizing the non-Gaussianity of the component ${\bf{w}}^T{\hat{\bf{x}}}$, where 
${\bf{w}}^T$ is a row of ${\bf{W}}$,
such that
the trasformed variables ${\bf{y}}= {\bf{W}}{\bf{x}}$  are the independent components.

In particular the \fstica\ algorithm operates with a neg-entropy approximation \citep{hyvarinen_oja_2000,hyvarinen_1999},
which can assume three different forms,  
depending on the regular non-quadratic function chosen in its expression:
$g(u) = u^3$, $g(u) = u~ {\rm exp}(-u^2)$ or $g(u) = {\rm tanh}(u)$, where $u={\bf{w}}^T{\hat{\bf{x}}}$. 
In the following we 
indicate these functions as $p$, $g$ and $t$ respectively.
The best choice of the function depends on the statistical proprieties of the components:  
kurtosis, or $p$, can be used for sub-Gaussian components in absence of outliers;  
$g$ may be better when the components are higly super-Gaussian or when robustness is important; 
$t$ is a general-purpose function \citep{hyvarinen_1999}. However we do not know a priori the 
statistics of the independent signals.   

Once the separation matrix ${\bf{W}}$ is obtained, 
since we have ${\bf{x}} = {\bf{W}^{-1}}{\bf{y}}$, 
we can derive the frequency 
scalings for each independent component:
the scaling between $\nu$ and $\nu '$ of the $j^{\rm th}$ component is given by the ratio
of ${W}^{-1}_{\nu j}/{W}^{-1}_{\nu ' j}$. 
If the spectral behaviour is given by a power law with index $\beta$, then we have
$\beta=\log[{W}^{-1}_{\nu j}/{W}^{-1}_{\nu ' j}]/\log(\nu/\nu')$. Therefore a negative value 
of frequency scaling indicates that the recovered component does not have a physical behaviour. 

Furthermore we can estimate the noise in the reconstructed maps. If we perform noise constrained
realizations  ${\bf{n_x}}$ for each frequency channel, the corresponding noise realizations in the 
\fstica\ outputs  are given by ${\bf{W}}{\bf{n_x}}$.

\section{Application to BEAST data}
\label{beast}

\begin{table*}
\begin{center}
\caption{ICA frequency scaling of the astrophysical component from 
BEAST maps smoothed at the three angular resolution, 
and for the three galactic cuts.} 
\begin{tabular}{ccccccccccccc}
\hline\hline
Resolution && \multicolumn{3}{c}{$|b|\leq17.5^{\circ}$} &&
 \multicolumn{3}{c}{$|b|\leq20^{\circ}$} &&
 \multicolumn{3}{c}{$|b|\leq22^{\circ}$}\\
 && $p$ & $g$ & $t$ && $p$ & $g$ & $t$ && $p$ & $g$ & $t$ \\
\hline
$30'$ && 0.221 & 0.129 & 2.382 && 0.190 & 0.118 & 2.470 && 0.177 & 0.068 & 2.480 \\
$40'$ && 0.127 & 1.296 & 2.683 && 1.459 & 1.116 & 3.036 && 1.580 & 1.090 & 2.838 \\
$60'$ && 1.049 & 1.148 & 2.917 && 0.927 & ---$^{\ast}$ & 3.141 && 0.930 & 0.815 & 2.928 \\
\hline\hline
\multicolumn{13}{r}{$^{\ast}$ no convergence of \fstica\ algorithm}
\end{tabular}
 \label{scalingB}
\end{center}
\end{table*}
\begin{figure*}
\begin{center}
\includegraphics[angle=0,width=5.8cm]{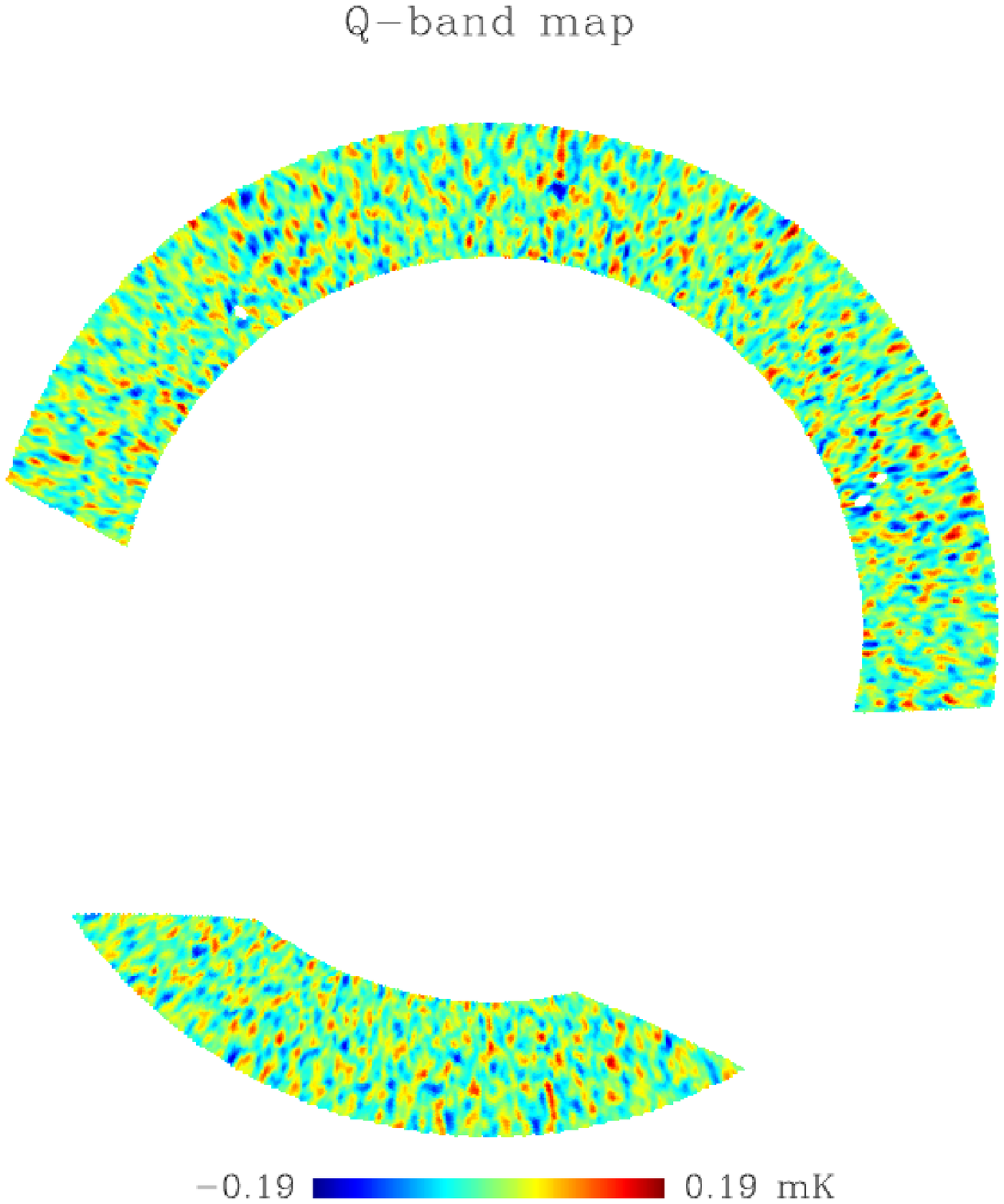}
\includegraphics[angle=0,width=5.8cm]{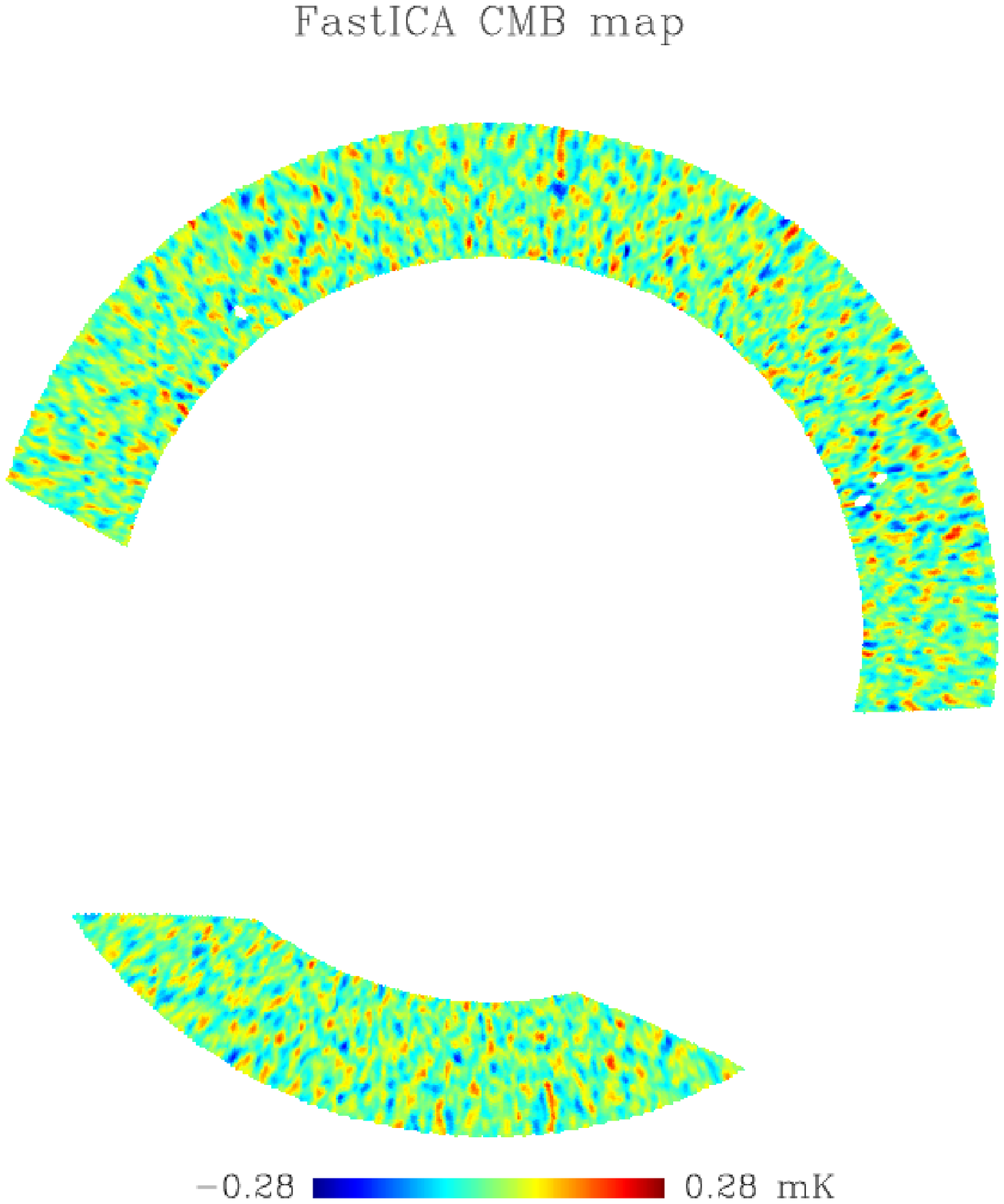}
\caption{left: The BEAST Q-band map smoothed to $40'$; right: 
The reconstructed astrophysical component (with $g$ function). 
The galactic plane is removed for $|b|\leq17.5^{\circ}$.} \label{icabeast}
\end{center}
\end{figure*}

The Background Emission Anisotropy Scanning Telescope (BEAST) is a 2.2 meter off-axis
Gregorian telescope \citep{figueiredo_etal_2005} with a focal plane consisting of 6 Q-band (38-45~GHz) 
and 2 Ka-band (26-36~GHz) corrugated scalar feed horns coupled with cryogenic 
HEMT amplifiers \citep{childers_etal_2005}. 
The instrument was installed at the UC White Mountain Research
Station at an altitude of 3.8 km in 2001 July. Data considered
here come from two different campaigns: one until December 2001 and
the second one in 2002 (February and August/September).

BEAST produced two maps covering an annular sky region around the NCP from 
$33^{\circ}<\delta<42^{\circ}$ 
with a resolution of $23'$ in Q-band and $30'$ in Ka-band. 
The sky maps are pixelized according to the
HEALPix\footnote{See http://www.eso.org/science/healpix/.} scheme
\citep{gorski_etal_1999} with a resolution parameter 
$N_{side}=512$ corresponding to pixel size of $6.9'$.

For a proper analysis of the results that \fstica\ will obtain,
we need to know
the instrumental noise properties, namely noise level
in the two frequency bands and its spatial distribution.
To estimate instrumental noise \citealp{meinhold_etal_2005}
have  made  ``difference'' maps at the two frequency bands separately. 
They binned data from first half of observation
into one map and the second half into another one. 
For each band the ``difference'' map is the pixel by pixel difference 
of these two maps, divided by 2 to maintain noise statistics as in the
sum map. Therefore these maps should not contain in principle, any sky signal, 
but only noise and the root mean square (rms)
of these maps is a measure of instrumental noise. The S/N ratio 
is quite poor: for the Q-band map \citealp{meinhold_etal_2005}
have found a value $\sim0.11$ at $23'$ resolution, which becomes $\sim0.57$ 
when the map is smoothed at $30'$. The Ka-band map shows a higher noise contribution. 
In addition from the ``difference'' maps we can see that noise is gaussian 
but, due to the scanning strategy, is not uniformly distributed on the sky.

Before applying \fstica\ a smoothing of the maps is required, due to the different 
angular resolution in the two bands, 
since in \fstica\ approach it is assumed a frequency-independent beam 
pattern (see Eq.~\ref{data_model}). 
We have smoothed the Ka and the Q maps to 
the same angular resolution, choosing values of $30'$, $40'$ and $60'$
in order to increase the S/N ratio. To obtain maps with significant 
signal level we decided not to smooth to resolution greater than $60'$, 
because of the 10~Hz high-pass filter applied to the BEAST data in the 
reduction processing: this indeed   
produces a signal cutoff on angular scales $\gsim6^{\circ}$.

We applied \fstica\ to the BEAST maps at the three resolutions working with 
all the non-quadratic functions described before identified by $p$, $g$ and $t$ respectively.
With two input sky maps \fstica\ is able to reconstruct only two outputs, since
the estimated matrix ${\bf{W}}$ has dimension $2\times2$. 
For all the considered cases one of the two is 
an astrophysical component, while the second component, due to the low S/N ratio, 
always has a negative frequency scaling, 
which suggests unphysical behaviour as explained in the previous section.
Performing a component separation on the full observed sky \fstica\ recovers an 
astrophysical signal with a frequency scaling consistent with free-free emission. 
This is due to the strong galactic emission 
in the two plane crossings.  
Therefore, in order to reconstruct the CMB component, we cut the sky regions 
where foreground emission overcomes CMB. 
We remove the three strongest point sources and the galactic plane, cutting out data 
within three different values of galactic latitude: $|b|\leq17.5^{\circ}$, 
$|b|\leq20^{\circ}$ and $|b|\leq22^{\circ}$. \citet{mejia_etal_2005} have   
estimated that, removing from BEAST maps regions with $|b|\leq17.5^{\circ}$, 
the individual galactic contributions remain below $\sim1\%$ of the map rms.

\subsection{The CMB component}
\label{results}

\begin{figure*}
\begin{centering}
\includegraphics[angle=90,width=0.7\textwidth]{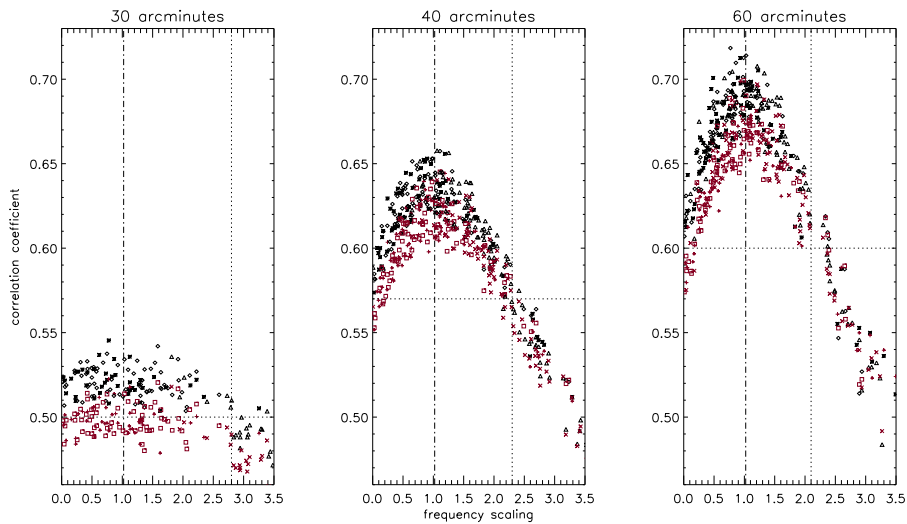}
\caption{Correlation between ICA CMB and simulated CMB in Ka (plus sign = $p$, square = $g$, `x' = $t$) and Q band (asterisk = $p$, diamond = $g$, triangle = $t$) towards 
recovered frequency scaling at $30'$, $40'$ and
$60'$. Line-dot lines show the expected scaling (1.022). Dot lines indicate the minimun correlation and the 
maximum scaling for a good CMB reconstruction. We can also observe that correlation with CMB-Q map is
greater than with CMB-Ka map.}
\label{rcorr_cfr}
\end{centering}
\end{figure*}

We wanted to verify if the component extracted by \fstica\ is indeed consistent with
CMB signal and the first figure of merit is the expected frequency 
scaling of the CMB between the two BEAST frequencies.
BEAST data are in antenna temperature and CMB fluctuations $\delta T_{A,{\rm CMB}}$ are related 
to brightness temperature fluctuation
$\delta T_{\rm CMB}$ by:
\beq \label{CMBscaling}
\delta T_{A, {\rm CMB}}(\nu)= \frac{x^{2}e^{x}}{(e^{x}-1)^{2}}\,
\delta T_{\rm CMB} \,,
\eeq
where $x=h\nu/kT_{\rm CMB}$. Assuming for the CMB a 
blackbody temperature $T_{\rm CMB}=2.725$ K \citep{mather_etal_1999},
we expect for the CMB component a frequency scaling between 
30 (Ka) and 41.5 (Q) GHz equal to 1.022. 

After removing the sky regions strongly contaminated by galactic emission, 
\fstica\ recovers an astrophysical component with frequency scaling quite different 
from that expected for CMB, as shown in 
Table~\ref{scalingB}.  
This is due to the high noise contribution, which affects also the recontructed 
astrophysical signal. Furthermore from Table~\ref{scalingB} we can observe that 
results obtained with $t$ function are in general worse than $p$ and $g$ results. 
Finally, increasing angular scale, with $p$ and $g$  the frequency scaling 
approaches the expected value in most of the cases. 
Nevertheless the spatial pattern of the recontructed astrophysical component 
resembles the Q map pattern, as shown in Fig.~\ref{icabeast} for the $40'$ smoothing case. 

After this first indication we proceeded by
verifying that reconstructed maps at different resolution 
are consistent one each other. 
After smoothing all maps down to $60'$, we have calculated the correlation 
between maps with different original resolution, finding Spearman correlation 
coefficients $r_{s}>0.8$ (with the exception of correlation between maps $30'-60'$ obtained with $t$   
function, for which $r_{s}\sim0.7$). 
This correlation indicates that \fstica\  recovers the same astrophysical signal  
at every resolution. Furthermore we observed that all the reconstructed maps have 
high spatial correlation with the Q band map smoothed at the same resolution, 
with $r_{s}>0.9$ except for $t$ results at $40'$ ($r_{s}\sim0.7$) and at $60'$ ($r_{s}\sim0.6$).
This is indeed expected since 
the S/N ratio is larger in Q band than in Ka. 
Finally, we verified that there are not significant changes in the astrophysical component 
reconstruction when extending the galactic cut. Indeed
reconstructed maps applying different cuts are consistent one each other 
($r_{s}\sim0.9$). Also the frequency scaling does not change   
significantly with the galactic cut (see Table~\ref{scalingB}) 
and there is not a well defined trend in the scaling variations with the cut extension.
This points out that the galactic contribution in the recontructed astrophysical component is not relevant.
Therefore we are confident that \fstica\ recovers a signal dominated by CMB anisotropies. 

\subsection{Testing results with Monte Carlo simulations}
\label{MC}

Given the poor S/N ratio in the BEAST data, we prefer to test the performance 
of \fstica\ by simulation. In order to analyse CMB reconstruction quality we performed 100 Monte Carlo simulations 
in which sky signal is simulated, observed following BEAST 
observing strategy, reduced as the actual data and then analysed by
\fstica.
We already demonstrated that foreground contribution is neglegible and therefore
we decided not to add any foreground templates.
CMB sky is generated according to the 
\WMAP\ best-fit power spectrum \citep{bennett_etal_2003} and 
convolved with a symmetric Gaussian beam with the BEAST angular resolution: 
$30'$ in Ka band and $23'$ in Q band. Maps are pixelized in HEALPix format 
\citep{gorski_etal_1999} with $N_{side}=512$.
Observing these maps, we produced Time Ordered Data (TOD) for each
BEAST channel from which we created maps in Ka and Q bands 
following the same reduction processing of the real BEAST data 
(see \citet{meinhold_etal_2005} for map-making process details).

As for instrumental noise simulation, we did not produce noise time streams for
each of BEAST detectors but we adopted a different recipe in order to have
noise maps with the same statistical properties as the actual data.
 
We generated white noise realizations with the same rms per pixel of BEAST maps 
in the two bands. It is clear that this is not enough due to the non-negligible
level of $1/f$ noise ($e.g.$ \citealt{meinhold_etal_2005}) which makes the noise
clearly not-white. We made use of the ``difference'' maps as derived by 
\citet{meinhold_etal_2005} to extract the noise angular distribution.
We then expanded in spherical harmonics both the white noise and the
``difference'' maps obtaining the harmonic coefficients $a_{\ell m}^w$ 
and $a_{\ell m}^d$ respectively. These coefficients are combined 
according to:
\beq
a_{\ell m}^{s}=\sqrt{\frac{\sum_{m}|a_{\ell m}^{d}|^{2}}
{\sum_{m}|a_{\ell m}^{w}|^{2}}}\,a_{\ell m}^{w}\,,
\eeq 
and we then generated noise maps with these new $a_{\ell m}^s$ coefficients. 
In this way noise simulated maps have also the same angular power spectra 
as the actual processed BEAST maps. We repeated this procedure for each
of the two BEAST frequency bands.
\begin{figure*}
\begin{centering}
\includegraphics[angle=90,width=0.8\textwidth]{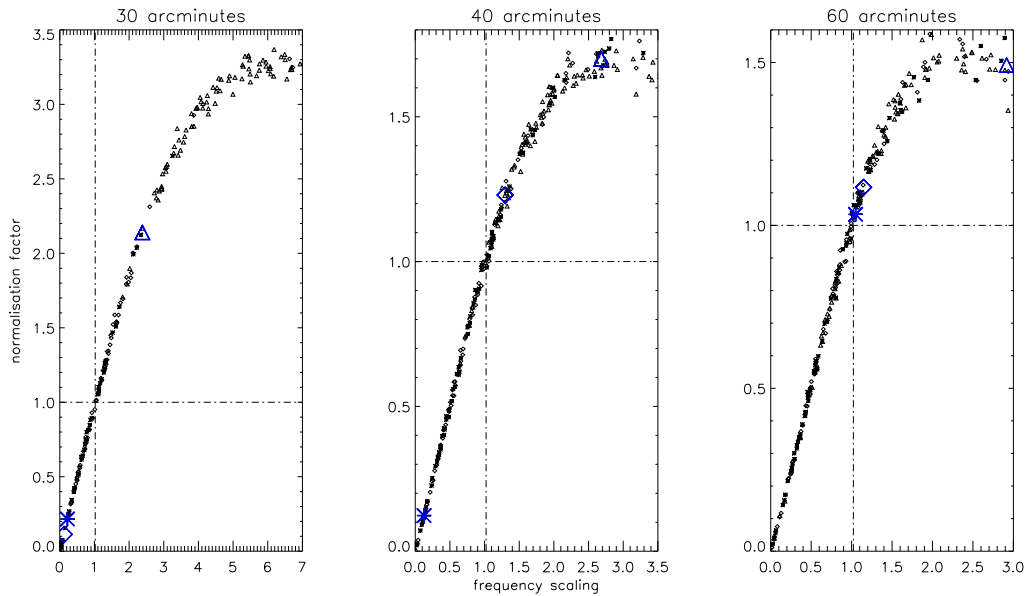}
\caption{Normalisation factor as a function of the frequency scaling for $30'$, $40'$ and $60'$. 
Line-dot lines indicate normalisation factor 
equal to 1 and the expected frequency scaling. The bigger symbols show the interpolation 
with the frequency scalings 
of the CMB recovered from BEAST data, cutting the galactic plane for $|b|\leq17.5^{\circ}$ 
and using $p$ (asterisks), 
$g$ (diamonds) and $t$ (triangles) function.}
\label{norm_inter}
\end{centering}
\end{figure*}
\begin{table}%[h]
\begin{center}
\caption{Simulated maps S/N ratios}\label{SN}
\begin{tabular}{ccccc}
\hline\hline
S/N && Ka & Q \\
\hline
$30'$ && $\sim0.09$ & $\sim0.57$ \\
$40'$ && $\sim0.28$ & $\sim0.81$ \\
$60'$ && $\sim0.33$ & $\sim0.85$ \\
\hline\hline
\end{tabular}
\end{center}
\end{table}
Finally, after smoothing to $30'$, $40'$ and $60'$ angular resolution, 
we added CMB and noise simulated maps together at each frequency, obtaining
simulated BEAST maps. 
Sub-pixel noise effects are negligible due to the high angular resolution, and the 
smoothing reduces any possible residual effect. 
possible 
In Table~\ref{SN}
we report the S/N ratios of the simulated maps. 
We underline that simulated Q maps smoothed to $30'$ have the 
same S/N ratio estimated by \citet{meinhold_etal_2005} for the BEAST $30'$ Q map 
derived with full processing of the data. This fact is a direct verification of
the success of our recipe for noise simulations.

Subsequently we applied \fstica\ to the simulated maps, after removing the region with $|b|\leq17.5^{\circ}$
as done for the actual data.
For every run we derived correlation coefficients of both
ICA maps with the input CMB.  
Therefore the ICA map of the two with the higher correlation coefficient 
is a possible CMB reconstruction. 

Figure~\ref{rcorr_cfr} shows the relation between correlation coefficients
and frequency scaling of the possible CMB reconstruction at the three different
angular resolutions for the three non-quadratic forms assumed by ICA.
It is interesting to note that in corrispondence of the expected CMB frequency
scaling (1.022), we observe the higher correlation coefficients and then
the best recovered CMB. We therefore use frequency scaling as a figure
of merit for the CMB reconstruction.
Furthermore we observe from Fig.~\ref{rcorr_cfr} that increasing angular resolution,
increases also values of the correlation: CMB reconstruction becomes better. This 
is expected because the S/N ratio increase with angular resolution (see Table~\ref{SN}).

Finally we used this relationship between correlation and frequency scaling to establish 
which CMB reconstructions are reliable. For every resolution we selected the minimum correlation 
value that characterizes a good reconstruction. This correlation 
coefficient $r_{s}$ corresponds to a maximum value of frequency scaling $s$. 
In Table~\ref{cmb_ok} we report the chosen values of 
correlation coefficient and frequency scaling (selecting regions near
the peaks in Fig.~\ref{rcorr_cfr}) and the number of ``good'' recovered CMB maps. 
As already noted in previous works (e.g. \citealt{maino_etal_2002, maino_etal_2003})
\fstica\ results with $t$ function are the worst, while $g$ performs better in 
the astrophysical context. Increasing angular scale increases the number of 
``good'' CMB reconstructions and decreases the differences between $p$, $g$ and $t$.  

\begin{table}
\begin{center}
\caption{Number of ``good'' CMB reconstructions from 100 simulated maps 
for each angular scale using  $p$, $g$ and $t$ function in \fstica\ algorithm.
See text for explanation.}
\label{cmb_ok}
\begin{tabular}{ccccc}
\hline\hline
$30'$ & $p$ & $g$ & $t$ \\
 $r_{s}>0.50$ & 43 & 70 & $\,$8 \\
 $s<2.8$ & 42 & 70 & $\,$6 \\
\hline
$40'$ & $p$ & $g$ & $t$ \\
 $r_{s}>0.57$ & 69 & 82 & 70 \\
 $s<2.3$ & 69 & 82 & 71 \\
\hline
$60'$ & $p$ & $g$ & $t$\\
 $r_{s}>0.60$ & 76 & 82 & 76 \\
 $s<2.1$ & 76 & 79 & 74 \\
\hline\hline
\end{tabular}
\end{center}
\end{table}

Looking at the CMB recovered from the data, we verify that the frequency scalings
reported in Table~\ref{scalingB} lie within the range of values that identify at every 
resolution a ``good'' reconstruction, with the exception of $t$ results at $40'$ and $60'$. 
This is a further indication of the fact that \fstica\ is able to extract reliable
CMB signal.

\section{normalisation of the CMB component}
\label{normal}

In general \fstica\ recovers a copy of the original signal \ie\ it is not able, in principle,
to recover the variance of the underlying sources. 
Therefore we have to normalize the CMB component recovered from BEAST data.
Generally we can derive the right normalisation factor directly from the \fstica\ outputs, but in this 
case, due to the poor S/N ratio, 
we must again use our Monte-Carlo 
simulations. In fact for each CMB recovered from simulations the scale factor 
is just the ratio between reconstructed and input CMB maps rms. 
The output CMB map is in Ka band antenna temperature and we 
compare it with the simulated CMB map in Ka band.
However noise in the reconstructed CMB is quite important and we
have to estimate and subtract it (at least in terms of rms).
Although the \fstica\ algorithm is higly non-linear, the data model is
linear \ie\ sources are obtained with a linear combination of the
input data. In this way we can quite easily estimate the noise
contribution in the reconstructed components by exploiting the
separation matrix $\mathbf{W}$ elements pertinent to the
CMB component.
We then subtracted, for each simulation, noise rms from the the output CMB rms to obtain
the exact recovered CMB rms to compare with the input CMB rms. 

Results are shown in Fig.~\ref{norm_inter} where normalisation factor is reported as function
of the frequency scaling at the three different angular resolutions.
There is a clear relationship between normalisation factor and frequency scaling. Indeed this relation
is almost linear within a frequency scaling $s_{max}$ and a normalisation factor $N_{max}$. 
Such values decrease when angular scale increases since \fstica\ performs better
(see Table~\ref{cmb_ok}) and scalings identifying a ``good'' CMB reconstruction 
are within $s_{max}$. Furthermore when the CMB component has the expected frequency scaling, 
it has also the right normalisation, with scale factor equal to 1. 

In Figure~\ref{norm_inter} we also report the frequency scalings obtained from BEAST data
at the three angular resolutions and for the three non-quadratic functions. We derived the
normalisation factor by simple linear interpolation of this relation in the points corresponding
to the actual frequency scalings.

\begin{figure}
\begin{centering}
\includegraphics[angle=90,width=0.4\textwidth]{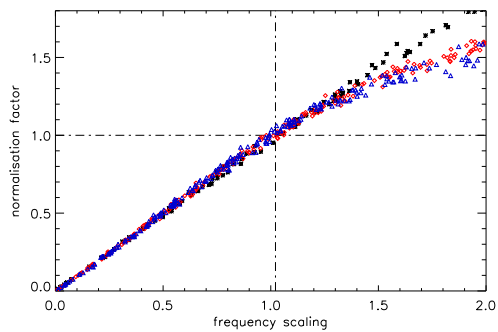}
\caption{Normalisation factor as a function of frequency scaling for $30'$ (asterisks), $40'$ (diamonds) 
and $60'$ (triangles).}
\label{norm_cfr}
\end{centering}
\end{figure}

Finally we observe that this relation
does not depend on the functions $p$, $g$ and $t$, and not even on angular resolution for 
scaling smaller than the theoretical one as shown in Fig.~\ref{norm_cfr}.

\section{Power spectrum}
\label{spectrum}

We estimated the angular power spectrum of the \fstica\ CMB component from
BEAST data and compared it with that derived by \citet{odwyer_etal_2005} for 
the analysis of Q band BEAST data. We extracted the spectrum choosing an 
angular resolution of $40'$. This 
is a good  
compromise between S/N ratio and signal level which is affected by the
10 Hz high-pass filter applied to the data. We considered the more conservative 
galactic cut ($|b|\leq17.5^{\circ}$) and used the $g$ function in the ICA 
algorithm which has the better recovered frequency scaling (see Table~\ref{scalingB}). 

To extract the CMB power spectra we adopted the MASTER method
(Hivon \etal~2002) that was also used by \citet{odwyer_etal_2005}.
MASTER returns a binned pseudo-$C_{\ell}$ estimator 
allowing for 
de-biasing the power spectrum for the effects specific 
of the experimental CMB observation, such as sky-cut, scanning strategy, 
data processing and instrumental noise.
This is expressed by the following data model:
\beq
\tilde{C}_\ell = \sum_{\ell'} M_{\ell \ell'} F_{\ell'} B^2_{\ell '}C_{\ell '}^{\rm th} + \langle N_\ell \rangle\, ,
\label{master}
\eeq 
where $\tilde{C}_\ell$ is the observed power spectrum, $C_\ell^{\rm th}$ is the
theoretical one. The $B^2_{\ell}$ term includes both
instrumental and pixel window functions and the kernel $M_{\ell \ell'}$ accounts 
for the mode-mode coupling between different modes due to the incomplete sky coverage and depends 
on the actual shape of the observed sky region (so it can be computed once for all).
The other terms are calibrated against Monte Carlo simulations.
In particular with simulations of CMB only observations we
compute the instrumental transfer function $F_{\ell}$, which accounts for data processing effects. 
Instrumental noise only simulations are needed to estimate the average noise
angular power spectrum $\langle N_\ell \rangle$, while
from simulated skies (CMB + noise) we derive errors on our final power spectrum estimation  
(see \citet{hivon_etal_2002} for details).
 
For a proper application of MASTER to a \fstica\ CMB map,
we must take into account that an ICA CMB map is indeed a linear combination of two 
data maps: the one in Ka-band and the one in Q-band,
\ie\ ${\bf y}_{\rm CMB}=w_{Ka}{\bf x}_{Ka}+w_{Q}{\bf x}_{Q}$,
where the weights $w_{Ka}$ and $w_{Q}$ are derived from the ICA separation matrix $\bf{W}$. 
In general, due to the different S/N ratios, we have obtained $w_{Q}>w_{Ka}$
(for example in the chosen case we have $w_{Ka}\sim0.20$ and $w_{Q}\sim1.04$), 
however the ICA CMB power spectrum is affected by experimental observation effects in both the bands.
In order to evaluate such effects we made use of the Monte-Carlo simulations 
already performed at $40'$. First we used simulated Ka-band and Q-band CMB 
to estimate the instrumental transfer functions for Ka and Q bands separately,   
$F_{\ell}^{Ka}$ and $F_{\ell}^{Q}$ respectively. We then computed the final 
ICA transfer function as $F_{\ell}=w_{Ka}^{2}F_{\ell}^{Ka}+w_{Q}^{2}F_{\ell}^{Q}$.  
Intrumental noise, and its angular power spectrum, in the reconstructed 
CMB component is obtained  in a similar manner: we use the same 
coefficients $w_{Ka}$ and $w_{Q}$ to properly combine the Ka-band and 
Q-band instrumental noise realizations.
Finally the same weights are used in signal plus noise simulations
in order to derive final error on the MASTER power spectrum.

We extracted the binned power spectrum 
choosing the same multipole bins used by \citet{odwyer_etal_2005}, 
with $\Delta\ell=55$. We estimated the 
ICA power spectrum for multipoles $\lsim400$ since signal 
is significantly suppressed at higher multipoles due to the 
selected $40'$ smoothing.

\begin{figure*}
\begin{centering}
\includegraphics[angle=90,width=0.7\textwidth]{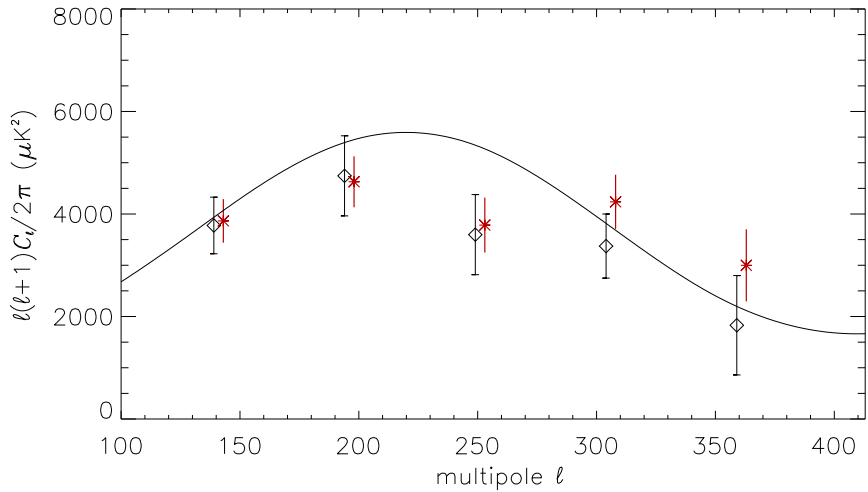}
\caption{CMB power spectra from BEAST data: stars show the spectrum extracted from \fstica\ $40'$ 
CMB map, diamonds show the $23'$ Q band map spectrum \citep{odwyer_etal_2005}.
ICA spectrum is shifted by $\Delta\ell=4$ for clarity. Solid line is the best-fit $WMAP$ power spectrum.}
\label{psbeast}
\end{centering}
\end{figure*}
\begin{table*}
\begin{center}
\caption{ The BEAST $C_{\ell}$ and $1\sigma$ error values (in $\mu{\rm K}^{2}$).}
\begin{tabular}{ccccccc }
\hline\hline
bin && \multicolumn{2}{c}{\fstica\ map} && \multicolumn{2}{c}{Q map}\\
$\ell_{min}-\ell_{max}$ && $\ell(\ell+1)C_{\ell}/2\pi$  & $1\sigma$ error && $\ell(\ell+1)C_{\ell}/2\pi$  & $1\sigma$ error\\
\hline
139-193 && 3865 & $\pm$425 && 3776 & $\pm$552 \\
194-248 && 4628 & $\pm$497 && 4744 & $\pm$781 \\
249-303 && 3783 & $\pm$535 && 3597 & $\pm$782 \\
304-358 && 4237 & $\pm$528 && 3374 & $\pm$625 \\
359-413 && 2997 & $\pm$703 && 1829 & $\pm$969 \\
\hline \hline
\end{tabular}
\label{valori_ps}
\end{center}
\end{table*}

Finally, since \fstica\ does not recover a CMB signal with the right variance, 
we normalised the spectrum using the scale factor derived from interpolation  
of the ``normalisation factor - frequency scaling'' relation,
as described in the previous Section (see Fig.~\ref{norm_inter}).

The resulting $40'$ \fstica\ CMB power spectrum is shown in Fig.~\ref{psbeast}, 
compared with the $23'$ Q-band map power spectrum estimated by  
\citet{odwyer_etal_2005} and with the best-fit 
\WMAP\ model. The agreement between the spectra is good. 
In particular the two BEAST spectra agree within $1-\sigma$. 
In Table~\ref{valori_ps} we report the 
$C_{\ell}$ and associated $1-\sigma$ errors for the two BEAST power spectra. 
Furthermore, both from Fig.~\ref{psbeast} and Table~\ref{valori_ps}, we can 
observe that ICA spectrum error bars are smaller than those of Q-band power spectrum. 
This is due to the smoothing at $40'$, which reduces the noise contribution.

The spectra agreement is a strong indication of the suitability of the 
\fstica\ CMB signal, and also of the goodness of the adopted normalisation procedure.

\section{Critical Discussion and Conclusion}
\label{conclusion}

In this paper we applied \fstica\ algorithm to real CMB data from the 
BEAST experiment. This is a ground-based experiment operating from the
UC White Mountain Research Station (CA) at an altitude of 3.8 km that
produced partial sky maps in two frequency bands (Ka and Q) with 
angular resolution of 30$'$ and 23$'$ respectively.

One of the \fstica\ requirements is that the instrumental noise has
to be Gaussian and uniformely distributed on the sky. This is
not the case for BEAST, which clearly shows $1/f$ noise and non-uniform 
integration time due to the 
observing strategy. The $1/f$ noise has been
accounted for by applying a high-pass filter to the Time
Ordered Data in the map-making process. This of course
alleviate the impact of non-white noise but also reduces
sky signal on large angular scales. Furthermore the signal-to-noise
ratio, as estimated from ``difference'' maps by \citet{meinhold_etal_2005},
is quite poor being $\sim 0.11$ for Q-band and even lower in Ka-band.

Another limitation for \fstica\ applicability is that different frequency
channels have to be at the same angular resolution. This forces us to
further convolve our data set. We choose three different values for
resolution: 30$'$, 40$'$ and 60$'$. We did not apply a more aggressive
smoothing since the high-pass filter effect on data is a clear suppression
of signal on larger scales. Smoothing data allows us to reach
a slightly better S/N ratio which helps in the application of \fstica. 

All these constraints have the consequence that \fstica\ always extracts from BEAST data, 
one physical component while the other is clearly noise related.
Furthermore in order to extract a CMB component we have to cut out the galactic plane
where galactic emission is dominating over CMB since otherwise this would prevent us
from properly reconstructing CMB.
Nevertheless, after galactic cut, \fstica\ recovers 
a CMB-like component, but with frequency scalings (Table~\ref{scalingB}) 
quite different from the theoretical one. This is again due to 
the relatively high instrumental noise, which alters CMB reconstruction. Despite this 
first bad indication, with further analysis we verified that this component is indeed
dominated by CMB anisotropies.

In order to test our CMB results quality we ran \fstica\ on 100 Monte Carlo 
simulations of Ka- and Q-band data at the three selected angular resolutions. 
These have been simulated by creating fake CMB skies with angular power spectrum from the
best-fit model from $WMAP$ \citep{bennett_etal_2003}, observing these skies according to
BEAST scanning strategy and reducing data with the same pipeline applied to real data.
We finally superimposed instrumental noise with the same
statistical and spatial properties as the actual data.  
We derived
correlation coefficients between CMB ICA maps with the input CMB and
studied the relation between these coefficients and the recovered scaling frequency,
in order to use scaling as a figure of merit. 
For every resolution we selected the maximum scaling allowed 
for a ``good'' CMB reconstruction and the comparison of these values with the results 
out of BEAST data confirms that \fstica\ indeed recovers
a reliable CMB component. Furthermore the relation shows the increasing reconstruction 
quality with angular scale, since the increasing S/N ratio, 
and the different \fstica\ performance with the three non-quadratic functions, 
with the better and the worse results obtained using $g$ and $t$ functions respectively. 
This final indication agrees with previous works \citep{maino_etal_2002,maino_etal_2003}.

Since \fstica\ is not able to recover the variance of the independent components
(it recoves a ``copy'' of the independent underlying components), 
we again used Monte Carlo simulations to obtain a normalisation procedure 
for the CMB component. 
In fact in this case the scale factor is just the ratio 
between output and input CMB rms. For each resolution we found a clear relation between 
scale factor and frequency scaling that is almost linear within certain values of scaling and 
normalisation factor.
The decreasing of such values with angular scale indicates again the 
corresponding improvement of \fstica\ performance due to the better S/N ratio.  
Furthermore the relation does not depend on the non-quadratic function and 
shows that those CMB reconstructions with the expected frequency scaling have
also the correct normalisation ($e.g.$ equal to 1). 
The normalisation factors for \fstica\ results out of BEAST data are 
derived by interpolation of this relation at the derived frequency scalings.

Finally we extracted the \fstica\ CMB angular power spectrum adopting a 
MASTER approach \citep{hivon_etal_2002} and normalised it with the proper scale factor.
We found a very good agreement with our results and the best-fit \WMAP\ model and also 
with the spectrum estimated from the BEAST Q-band map \citep{odwyer_etal_2005}, although
on a limited multipole range because of the extra-smoothing applied to the data.
This spectra agreement confirms the reliability of the CMB extracted by \fstica\ and validates 
our normalisation procedure. 

Our analysis, together with that on DMR data performed by \citet{maino_etal_2003}, 
confirms the very good performance of blind algorithms like \fstica\ in extracting
a CMB component even from noisy data on a small patch of the sky like
BEAST ones. Therefore we think that blind algorithms are valid tools for present
and future CMB experiments providing information on the independent component
in the actual observed sky signal which could be used to feed much more
complex algorithm like Maximum Entropy Method. This is particularly relevant
for future CMB polarisation experiments where we will be forced to work
with low S/N ratios and where our knowledge of polarisation for foregrounds
is still poor \citep{stivoli_etal_2005}.

\section*{Acknowledgements}
It is a pleasure to thank E.~Salerno for useful discussion. 
TV and CAW acknowledge FAPESP support for
BEAST under grant 00/06770-2. TV was partially
supported by CNPq grant 305219/2004-9 and
CAW by grant 307433/2004-8.

\label{lastpage}
\end{document}